\documentclass[twocolumn]{aastex62}

\usepackage[varg]{txfonts}
\usepackage{bm}

\newcommand{\beq}{\begin{equation}}
\newcommand{\eeq}{\end{equation}}
\newcommand{\beqn}{\begin{eqnarray}}
\newcommand{\eeqn}{\end{eqnarray}}
\renewcommand{\leq}{\leqslant}

\newcommand{\eqref}[1]{(\ref{#1})}

\newcommand{\dfrac}[2]{ {\displaystyle\frac{#1}{#2}} }

\newcommand{\pfrac}[2]{ \biggl(\dfrac{#1}{#2}\biggr) }

\newcommand{\kB}{k_{\rm B}}

\newcommand{\brac}[1]{\langle #1 \rangle}

\newcommand{\eps}{\epsilon}
\newcommand{\Ecrite}{E_{{\rm crit},e}}
\newcommand{\Ecriti}{E_{{\rm crit},i}}

\defcitealias{OI15}{Paper I}

\shorttitle{The Generalized Nonlinear Ohm's Law}
\shortauthors{Okuzumi et al.}

\begin{document}
\title{The Generalized Nonlinear Ohm's Law: How a Strong Electric Field Influences Non-ideal MHD Effects in Dusty Protoplanetary Disks}

\author{Satoshi Okuzumi}
\affiliation{Department of Earth and Planetary Sciences, 
Tokyo Institute of Technology, Meguro-ku, Tokyo 152-8551, Japan}

\author{Shoji Mori}
\affiliation{Department of Earth and Planetary Sciences, 
Tokyo Institute of Technology, Meguro-ku, Tokyo 152-8551, Japan}
\affiliation{Department of Astronomy, The University of Tokyo, Bunkyo-ku, Tokyo 113-0033, Japan}

\author{Shu-ichiro Inutsuka}
\affiliation{Department of Physics, Nagoya University, Nagoya, Aichi 464-8602, Japan}

\correspondingauthor{Satoshi Okuzumi}
\email{okuzumi@eps.sci.titech.ac.jp}

\begin{abstract}
The magnetohydrodynamics (MHD) of protoplanetary disks are strongly subject to the non-ideal MHD effects arising from the low ionization fraction of the disk gas. A strong electric field induced by gas motions can heat ionized gas particles and can thereby affect the ionization balance in the disks. Our previous studies revealed that in dusty protoplanetary disks, the Ohmic conductivity decreases with increasing electric field strength until the electrical breakdown of the disk gas occurs. In this study, we extend our previous work to more general cases where both electric and magnetic fields affect the motion of plasma particles, allowing us to study the impacts of plasma heating on all non-ideal MHD effects: Ohmic, Hall, and ambipolar diffusion. We find that the upper limit on the electric current we previously derived applies even in the presence of magnetic fields. Although the Hall and ambipolar resistivities can either increase or decrease with electric field strength depending on the abundance of charged dust grains, the Ohmic resistivity always increases with electric field strength. An order-of-magnitude estimate suggests that a large-scale electric current generated by gas motions in the inner part of protoplanetary disks could exceed the upper limit. This implies that MHD motions of the inner disk, such as the motion driven by the Hall-shear instability, could either get suppressed or trigger electrical breakdown (lightning discharge). This may have important implications for gas accretion and chondrule formation in the inner part of protoplanetary disks.  
\end{abstract}
\keywords{accretion, accretion disks --- instabilities -- magnetohydrodynamics (MHD) --- planets and satellites: formation --- plasmas --- protoplanetary disks --- turbulence} 

\section{Introduction}\label{sec:intro}
The dynamics and evolution of protoplanetary disks is key to understanding 
formation occurring in the disks.
Like in many astrophysical systems, magnetic field is thought to play an important role
in the gas dynamics in the disks. 

However, because the interior of protoplanetary disks is poorly ionized, 
the MHD of the disks is strongly subject to the effects arising from the finite electrical 
conductivities of the gas, the so-called non-ideal MHD effects (see \citealt{T+14} for a review).
It has long been recognized that Ohmic diffusion stabilizes the magnetorotational instability 
(MRI; \citealt{BH91}) in the dense, cool part of the disks \citep[e.g.,][]{G96,SMUN00}.
However, Ohmic diffusion is the dominant non-ideal MHD effect 
only when the gas drag acting on the plasma particles is stronger than the 
magnetic Lorentz force acting on them, which is only the case in the inner part of the disks \citep{KB04,W07,B11a}. 
In fact, recent MHD simulations have shown that ambipolar diffusion, another non-ideal MHD effect, 
substantially damps the MRI on the surface and in the outer part of the disks \citep{BS13a,SBA+13a,
SBA+13b,GTNM15}. 
The Hall drift, the last non-ideal MHD effect, further changes the disk dynamics
by introducing the Hall-shear instability (HSI) to disks whose 
net vertical magnetic field has the same direction as the disk rotation vector \citep{K08,WS12}.
The Hall-shear instability amplifies horizontal magnetic fields, thus allowing for  
accretion in the relatively inner part of the disk \citep{B14,B15,B17,LKF14}. 
Thus, it is essential to fully understand the conductivity of gas in protoplanetary disks.

Conventionally, models of disk ionization in protoplanetary disks \citep[e.g.,][]{SMUN00,IN06a,W07,BG09,O09}
assume that the ionization state is determined by the balance between external ionization and recombination 
of plasma particles within the gas and on dust grains. 
It is also assumed implicitly that the electric field as measured in the comoving frame of the gas,
which drives electric current in the disk, is so weak that it has no effect on the plasma temperatures. 
However, because the gas disk is poorly conducting, the electric field accompanied by the MHD motions of the disks
can be strong, potentially affecting the plasma temperatures and even the ionization balance. 
\citet{IS05} first pointed out that the electric field inside MRI-driven turbulence could be 
strong enough to cause electrical breakdown of the disk gas, potentially leading to 
self-sustained turbulence in which the high electric conductivity provided 
by the breakdown keeps the MRI active without additional ionizing sources.
This self-sustaining mechanism for the MRI in protoplanetary disks was later confirmed by \citet{MOI12}, 
who performed MHD simulations in which the Ohmic conductivity was allowed to 
increase arbitrarily at a certain electric field strength to mimic electrical breakdown.

More recently, \citet[][henceforth \citetalias{OI15}]{OI15} 
developed a charge reaction model that incorporates plasma heating by electric field, 
showing that plasma heating does not only enhance but can also suppress
the conductivity of disk gas. 
They found that as plasma particles are electrically heated, they collide with and stick onto 
dust grains more frequently, and consequently the plasma densities in the gas decrease. 
The decrease in the conductivity with increasing electric field strength gives rise to 
an upper limit on the electric current that can only be exceeded with electrical breakdown.
This effect could quench MHD turbulence before the electric field strength reaches 
the breakdown threshold \citep{MO16,MMOI17}.
The upper limit on the plasma current can also facilitate charge separation 
and even lead to lightning discharge. 
\citep{JO18} recently propose that this lightning discharge might have 
led to the formation of chondrules, 
millimeter-sized solid particles that experienced melting by flash heating events, 
found in meteorites \citep[see, e.g.,][for the lightning scenario for chondrule formation]{W66,DC00}. 

There are two important limitations to the formulation of \citetalias{OI15}. 
Firstly, we assumed that collisions between electrons and neutrals
are purely elastic. This assumption is valid at low electric field strengths where 
electron heating starts to be effective, but breaks down 
once the electron energy exceeds the excitation energies of neutrals.   
By neglecting inelastic energy losses, we overestimated the electron 
energy at high electric field strengths. 
Secondly, we neglected the effects of magnetic fields on the motion of plasma particles
by assuming that the gas drag force acting on the particles is stronger than the magnetic Lorentz force.
For this reason, the previous model is not able to treat Hall drift and ambipolar diffusion, 
and therefore inapplicable to the dense inner part of the disks where these non-Ohmic effects dominate.
The purpose of this paper is to reformulate the work of \citetalias{OI15} and provide 
a model that can treat all non-ideal MHD effects as well as non-elastic plasma--neutral collisions. 

This paper is organized as follows. 
In Section~\ref{sec:kinetics}, we formulate the kinetics of charged gas particles 
in the presence of both electric and magnetic fields and illustrate how magnetic fields 
as well as inelastic energy losses affect the electric heating of plasmas.
This kinetic model is used in Sections~\ref{sec:J} to present analytic estimates 
of the electric current at high electric field strengths in the presence of magnetic fields. 
In Section~\ref{sec:case}, we combine the kinetic model with a simplified charge reaction model
to demonstrate how the electric current and magnetic resistivities depend on the magnitude and direction of an applied electric field. 
Implications for the MHD of protoplanetary disks are discussed in Section~\ref{sec:discussion},  
and a summary is presented in Section~\ref{sec:summary}.

\section{Kinetics of Weakly Ionized Plasmas in Electric and Magnetic Fields}\label{sec:kinetics}
We begin by studying how the presence of magnetic field affects 
electric heating of plasmas in a neutral gas. 
In \citetalias{OI15}, we calculated the electron kinetic energy and other related 
quantities directly from the exact velocity distribution function for electrons. 
Such an approach is rigorous, but is useful only when a simple and closed analytic expression 
for the velocity distribution is known.
The electron velocity distribution adopted in \citetalias{OI15}, the so-called Davydov distribution \citep{D35},
is only valid when the electron--neutral collisions are purely elastic and when no magnetic field is present.  
Unfortunately, there is no known exact expression for the electron velocity distribution for more general cases.

In this study, we use an alternative approach based on the moment formalism \citep{GZS80}.
In this approach, we approximate the velocity distribution functions of ions and electrons 
with a Maxwellian whose center is offset in the velocity space. 
This offset Maxwellian distribution is characterized by 
the mean velocity  and mean energy of the plasma particles,
which we determine by solving the first and second moment equations of the Boltzmann equation 
including the magnetic Lorentz force.
This way, we are able to compute the mean and random velocities of a plasma 
in the presence of both magnetic and electric field without having to solve the original Boltzmann equation numerically.   
In \citetalias{OI15}, we already applied this approach to ions, whose velocity distribution 
has no exact analytic expression even in the absence of magnetic field \citep{W53}.  
In this study, we apply this approach to both ions and electrons. 

\subsection{Moment Equations for the Drift Velocities and Mean Energies}\label{sec:moment}
We denote the velocity of plasma particles (ions and electrons) relative to the neutral gas by ${\bm v}_\alpha$ 
and their velocity distribution function by $f_\alpha({\bm v}_\alpha)$, where $\alpha=e$ for electrons
and $\alpha = i$ for ions. 
The mean drift velocity and mean energy, $\brac{{\bm v}_\alpha}$ and $\brac{\eps_\alpha}$, are defined as
\beq
\brac{{\bm v}_\alpha} \equiv \int {\bm v}_\alpha f_\alpha({\bm v}_\alpha) d^3 v_\alpha,
\label{eq:v_def}
\eeq
\beq
\brac{\eps_\alpha} \equiv 
\frac{m_\alpha}{2} \int v_\alpha^2 f_\alpha({\bm v}_\alpha) d^3 v_\alpha,
\label{eq:e_def}
\eeq
respectively, where $m_\alpha$ is the particle mass and $v_\alpha = |{\bm v}_\alpha|$.

We consider the motion of plasma particles in a neutral gas with electric and magnetic fields. 
If the plasma density is much lower than the neutral gas density, 
the motion is determined by the Lorentz force and 
drag force arising from the collisions with the neutrals. 
In the reference frame where the mean velocity of the neutrals vanishes, 
the first and second moments of the Boltzmann equation for plasma particles 
are given by \citep[][Chapters 6 and 9]{GZS80}
\beq
m_\alpha\frac{d\brac{{\bm v}_\alpha}}{dt} =  
q_\alpha\left( {\bm E}' + \frac{\brac{{\bm v}_\alpha}}{c}\times {\bm B} \right) 
- \mu_{\alpha n} \nu_{\alpha n}  \brac{{\bm v}_\alpha},
\label{eq:dvdt0}
\eeq
\beq
\frac{d\brac{\eps_\alpha}}{dt} 
=  q_\alpha{\bm E}'\cdot\brac{{\bm v}_{\alpha}}
-  {\kappa_{\alpha n}\nu_{\alpha n}}\left(\brac{\eps_\alpha}- \frac{3\kB T}{2} \right),
\label{eq:dedt0}
\eeq
respectively. 
Here, ${\bm E}'$ and ${\bm B}$ are the electric and magnetic fields as measured in the neutral-rest frame, respectively;  
$c$ is the speed of light; 
$q_\alpha$ is the particle charge; 
$\mu_{\alpha n} = m_\alpha m_n/(m_\alpha + m_n)$, 
$\nu_{\alpha n}$, and $\kappa_{\alpha n}$ are the reduced mass, mean collision frequency, 
and mean energy transfer efficiency for the collision with the neutrals, respectively;
$\kB$ is the Boltzmann constant; 
and $T$ is the temperature of the neutral gas. 
In the right-hand sides of Equations~\eqref{eq:dvdt0} and \eqref{eq:dedt0}, 
the first terms represent the mean Lorentz force and mean work done by the electric field, respectively,
whereas the second terms represent the mean momentum and energy losses due to the collisions with the neutrals.
The prime in ${\bm E}'$ emphasizes that it is the electric field in the neutral-rest frame; 
in the frame where the neutral gas has a mean velocity ${\bm u}$, 
the electric field is given by ${\bm E} = {\bm E}' - {\bm u} \times {\bm B}/c$.

\subsection{Offset Maxwell Distribution Function}\label{sec:Maxwellian}
As stated earlier, we approximate $f_\alpha({\bm v}_\alpha)$ by 
the offset Maxwell distribution (\citealt{H39}; \citetalias{OI15})
\beq
f_\alpha({\bm v}_\alpha) = \pfrac{m_\alpha}{2\pi \kB T_\alpha}^{3/2}
\exp\left(-\frac{m_\alpha({\bm v}_\alpha-\brac{{\bm v}_\alpha})^2}{2 \kB T_\alpha}\right).
\label{eq:falpha}
\eeq
Here, the temperature $T_\alpha$ measures the kinetic energy of random motion,
related to $\brac{{\bm v}_\alpha}$ and $\brac{\eps_\alpha}$ as  
\beq
\frac{3}{2}\kB T_\alpha \equiv \brac{\eps_\alpha} - \frac{1}{2}m_\alpha \brac{{\bm v}_\alpha}^2,
\label{eq:Talpha}
\eeq
It can be easily checked that $\brac{{\bm v}_\alpha}$ and $\brac{\eps_\alpha}$ 
in Equation~\eqref{eq:falpha} satisfy their definitions, Equations~\eqref{eq:v_def} and \eqref{eq:e_def}. 

For electrons, the velocity distribution in a weakly ionized gas tends to be nearly isotropic,
i.e., $|\brac{{\bm v}_e}| \ll \sqrt{\kB T_e/m_e}$, because of $m_e \ll m_n$ (see \citealt{GZS80}, Section~5.2). 
Therefore, we may approximate $f_e$ to first order in $\brac{{\bm v}_e}$
to obtain a simpler expression 
\beq
f_e({\bm v}_e) = \pfrac{m_e}{2\pi \kB  T_e}^{3/2}
\left(1+\frac{m_e{\bm v}_e\cdot\brac{{\bm v}_e}}{\kB T_e}\right)
\exp\left(-\frac{\eps_e^2}{\kB T_e}\right)
\label{eq:fe}
\eeq
with
\beq
\frac{3}{2}\kB T_e = \brac{\eps_e}.
\label{eq:Te}
\eeq
We will use these expressions for electrons instead of using Equations~\eqref{eq:falpha} and \eqref{eq:Talpha}. 
We will also use that, to zeroth order in $\brac{{\bm v}_e}$, 
the electron speed $v_e = |{\bm v}_e|$ has a mean value  
\beq
\brac{v_e} \approx \sqrt{\frac{8\kB T_e}{\pi m_e}} = 4\sqrt{\frac{\brac{\eps_e}}{3\pi m_e}}.
\label{eq:ve}
\eeq

\subsection{Collision Frequencies and Energy Transfer Efficiencies}
In principle, the collision frequencies $\nu_{\alpha n}$ and energy transfer efficiencies 
$\kappa_{\alpha n}$ are the sums of the contributions from elastic and inelastic collisions.
Elastic collisions conserve the kinetic energy of the relative motion between the colliding 
charged and neutral particles\footnote{Note, however, that 
the kinetic energy of the charged particles  as measured in the {\it neutral-rest} frame 
does decrease through elastic collisions. 
This is the reason why elastic collisions contribute to $\kappa_{\alpha n}$.
}, 
whereas inelastic collisions do not because of excitation and ionization energy losses. 

In practice, the contributions of inelastic collisions are negligible for ions 
because the cross section for the inelastic collisions is small as long as $\eps_i \la 10~\rm eV$ \citep[e.g.,][Chapter 2.9]{GZS80}.
Neglecting the inelastic collisions, $\nu_{in}$ and $\kappa_{in}$ are  given by \citep[e.g.,][Chapter 2]{GZS80}
\beq
\nu_{in} = K_{in} n_n,
\eeq
\beq
\kappa_{i n} = \frac{2m_i m_n}{(m_i+m_n)^2},
\label{eq:kappa_in}
\eeq
respectively.
Here, $K_{in}$ is the momentum transfer rate coefficient for (elastic) ion--neutral collisions,
which is independent of ion--neutral collision velocity because of 
the polarization force acting between them  \citep[see, e.g.,][]{W53}.
 We take $K_{in} = 1.6\times 10^{-9}~\rm cm^3~s^{-1}$ following \citet{NU86}.

For electrons, the contribution of inelastic collisions is negligible to $\nu_{en}$ 
(but not to $\kappa_{en}$ as we discuss later). 
Approximating $f_e$ by a Maxwellian with $\brac{{\bm v}_e} \approx 0$, 
$\nu_{en}$ can formally be written as \citep[][Section 6.3]{GZS80}
\beq
\nu_{en} = n_n \frac{\brac{\sigma_{en} v_e^3}}{\brac{v_e^2}} ,
\eeq
where $\sigma_{en}$ is the momentum transfer cross section for electron--neutral collisions.
Since $\sigma_{en}$ is almost constant ($\approx 10^{-15}~{\rm cm^{-3}}$) 
as long as $\eps_e \ll 10~{\rm eV}$ (see, e.g., Figure 5 of \citealt{YSH+08}),
we may approximate $\nu_{en}$ as  
\beq
\nu_{en} 
= \frac{4 n_n \sigma_{en}}{3}\brac{v_e}
= \frac{16 n_n \sigma_{en}}{3}\sqrt{ \frac{\brac{\eps_e}}{3\pi m_e} },
\label{eq:nu_en}
\eeq
where we have used that 
$\brac{v_e^3}/\brac{v_e^2} = 4\brac{v_e}/3$ under the Maxwellian approximation. 

In \citetalias{OI15}, we neglected inelastic contributions to $\kappa_{en}$ and used 
the expression for purely elastic collisions, $\kappa_{en} = {2m_e m_n}/{(m_e+m_n)^2} \approx {2m_e}/{m_n}$.
In reality, the inelastic contributions are not always negligible since the elastic contribution $\approx {2m_e}/{m_n}$ is also small. 
To account for the inelastic energy losses, we {express} $\kappa_{en}$ as 
\beq
\kappa_{en} =  \frac{1}{P_\ell}\frac{2m_e}{m_n},
\label{eq:kappa_en}
\eeq
where $P_\ell$ is a dimensionless factor depending on the mean electron kinetic energy $\brac{\eps_e}$.
The factor $P_\ell$ stands for the fractional contribution of elastic collisions to the total collisional energy loss of elections,
with $P_\ell = 1$ if the electron--neutral collisions are purely elastic and $P_\ell < 1$ otherwise.
Based on the results of theoretical calculations for electron--$\rm H_{\rm 2}$ collisions (see Appendix \ref{sec:Pell}),
we evaluate $P_\ell$ as  
\beq
P_\ell = \left(1+\frac{\brac{\eps_e}}{0.0075~{\rm eV}}\right)^{-1/2}.
\label{eq:Pell_model}
\eeq 
{This expression reproduces theoretical estimates for $P_\ell$ to within a factor of 2 
as long as $\kB T \la 5~\rm eV$; for higher electron energies, \eqref{eq:Pell_model} could underestimate energy losses due to electronic excitation and ionization (see Appendix \ref{sec:Pell} for details).} 
According to Equation~\eqref{eq:Pell_model}, the collisions can be regarded as elastic ($P_\ell \approx 1$) only when $\brac{\eps_e}  \ll 0.01~{\rm eV}$.

\subsection{Steady State Solution and the Effective and Critical Field Strengths}\label{sec:steady}
Since plasma particles in protoplanetary disks frequently collide with neutrals, 
we may assume that the velocity distribution functions for the plasmas are in steady state on the dynamical timescale of the disks. 
Below we derive the steady solutions of the moment equations.

Equation~\eqref{eq:dvdt0} is the basic equation for the standard generalized Ohm's law,
and its steady-state solution is already known to be \citep[e.g.,][]{NU86}
\beq
\brac{{\bm v_\alpha}}
= \frac{q_\alpha}{\mu_{\alpha n}\nu_{\alpha n}}
\left( 
{\bm E}'_\parallel - \dfrac{\beta_\alpha}{1+\beta_\alpha^2}\hat{\bm B}\times {\bm E}'_\perp
+  \dfrac{{\bm E}'_\perp}{1+\beta_\alpha^2}  
\right),
\label{eq:valpha}
\eeq
where ${\bm E}'_\parallel$ and ${\bm E}'_\perp$ are the components of ${\bm E}'$ parallel and 
perpendicular to ${\bm B}$, respectively, $\hat{\bm B}$ is the unit vector of ${\bm B}$, 
and $\beta_\alpha$ is the Hall parameter defined by 
\beq
\beta_\alpha =  \frac{q_\alpha B}{\mu_{\alpha n}\nu_{\alpha n}c}
\eeq
with $B \equiv |{\bm B}|$. 
The magnitude of the Hall parameter 
measures the relative importance of the magnetic Lorentz force to the neutral drag force \citep[see, e.g.,][]{WN99}.
If $|\beta_\alpha| \ll 1$, the the magnetic Lorentz force is negligible, the regime considered in \citetalias{OI15}. 
In general, one has $|\beta_e|/\beta_i \approx 440(T_e/300~\rm K)^{-1/2}$, 
independent of $B$ and $n_n$ \citep{W07}. 

Substituting Equation~\eqref{eq:valpha} into Equation~\eqref{eq:dedt0} 
and taking ${d\brac{\eps_\alpha}}/{dt} =0$, obtain an equation that determines $\brac{\eps_\alpha}$ for steady state, 
\beq
\frac{q_\alpha^2}{\mu_{\alpha n}\nu_{\alpha n}}E_{{\rm eff},\alpha}'^2
-  {\kappa_{\alpha n}\nu_{\alpha n}}\left(\brac{\eps_\alpha}- \frac{3\kB T}{2} \right) = 0,
\label{eq:dedt_eff}
\eeq
where we have introduced the effective electric field strength for charged species $\alpha$
\citep[][Chapter 5]{GZS80}, 
\beq
E_{{\rm eff},\alpha}' \equiv \sqrt{E_\parallel'^2  + \frac{E_\perp'^2}{1+\beta_\alpha^2}},
\label{eq:Eeff}
\eeq
with $E_\parallel' \equiv |{\bm E}'_\parallel|$ and $E_\perp' \equiv |{\bm E}'_\perp|$. 
Equation~\eqref{eq:dedt_eff} is key to understanding how magnetic fields affect the electric heating of plasmas. 
It is $E_{{\rm eff},\alpha}'$, not the magnitude of the total electric field $E' = (E_\parallel'^2 + E_\perp'^2)^{1/2}$, 
that determines the rate of plasma heating (the first term in Equation~\eqref{eq:dedt_eff}) in the presence of a magnetic field.  
Because $E_{{\rm eff},\alpha}' \leq E'$, magnetic fields generally suppress the plasma heating. 
In particular, when ${\bm E}' \perp {\bm B}$ (i.e., $E'_\parallel=0$ and $E'_\perp=E'$) and $|\beta_\alpha| \gg 1$, 
$E_{{\rm eff},\alpha}'$ is smaller than $E'$ by a factor of $1/|\beta_\alpha| \ll 1 $.

For ions, $\nu_{in}$ is independent of $\brac{\eps_i}$, and hence Equation~\eqref{eq:dedt_eff} can be analytically solved as
\beq
\brac{\eps_i} = \frac{3}{2}\kB T + \frac{e^2}{\mu_{in} \kappa_{in} \nu_{in}^2}E_{{\rm eff},i}'^2.
\label{eq:ei_general}
\eeq
In the right-hand side of Equation~\eqref{eq:ei_general}, 
the first term indicates that the ion temperature is equal to the 
neutral temperature in the absence of electric fields. 
The second term represents ion heating by the electric field, 
and is larger than the first term when $E_{{\rm eff},i} $ is above the threshold 
\beqn
\Ecriti &\equiv& \frac{\nu_{in}}{e} \sqrt{\frac{3\mu_{in}\kappa_{in}\kB T}{2}} 
\nonumber \\
&\approx& \frac{m_n n_n K_{in} }{e} \sqrt{\frac{3\kB T}{m_i}}.
\label{eq:Ecriti}
\eeqn
In the final expression, where we have used that $m_i \gg m_n$. 

For electrons, Equation~\eqref{eq:dedt_eff} with Equations~\eqref{eq:nu_en}--\eqref{eq:Pell_model} is a transcendental equation
for $\brac{\eps_e}$, which we must solve numerically.  
It is useful to note, however, that Equation~\eqref{eq:dedt_eff} can formally be rewritten as
\beq
\brac{\eps_e} = \frac{3}{2} \kB T\left( \frac{1}{2}+
\frac{1}{2}\sqrt{1+\frac{9\pi P_\ell}{16}\pfrac{E_{{\rm eff},e}'}{\Ecrite}^2}\right),
\label{eq:ee_general}
\eeq
where 
\beq
\Ecrite \equiv  
\sqrt{\frac{6m_e}{m_n}} \frac{n_n \sigma_{en} \kB T }{e}
\label{eq:Ecrit}
\eeq
is the critical field strength for electron heating introduced in \citetalias{OI15}.\footnote{Equation~\eqref{eq:ee_general} 
is identical to Equation (17) of \citet{MO16} except that the factor $2/3$ in the previous expression 
has now been replaced by ${9\pi P_\ell}/{64}$.
The new expression is more accurate because it takes into account inelastic energy losses, 
and because it uses the exact expression for $\nu_{en}$ (Equation~\eqref{eq:nu_en}) 
whereas the previous expression used an approximate expression $\nu_{en} = n_n \sigma_{en} \brac{v_e^2}^{1/2}$. 
}
Equation~\eqref{eq:ee_general} suggests that electron heating occurs at 
$E'_{{\rm eff},e} \ga E_{{\rm crit},e}/\sqrt{P_\ell}$. 
Since $P_\ell \sim 1$ for $T_e \sim T \sim 100~\rm K$, 
we may assume that electron heating occurs 
when $E'_{{\rm eff},e}$ exceeds $E_{{\rm crit},e}$. 
Note that the right-hand side of Equation~\eqref{eq:ee_general} 
depends on $\brac{\eps_e}$ through $P_\ell$ and $\beta_e \propto 1/\nu_{en}$.

\begin{figure}[t]
\centering
\resizebox{\hsize}{!}{\includegraphics{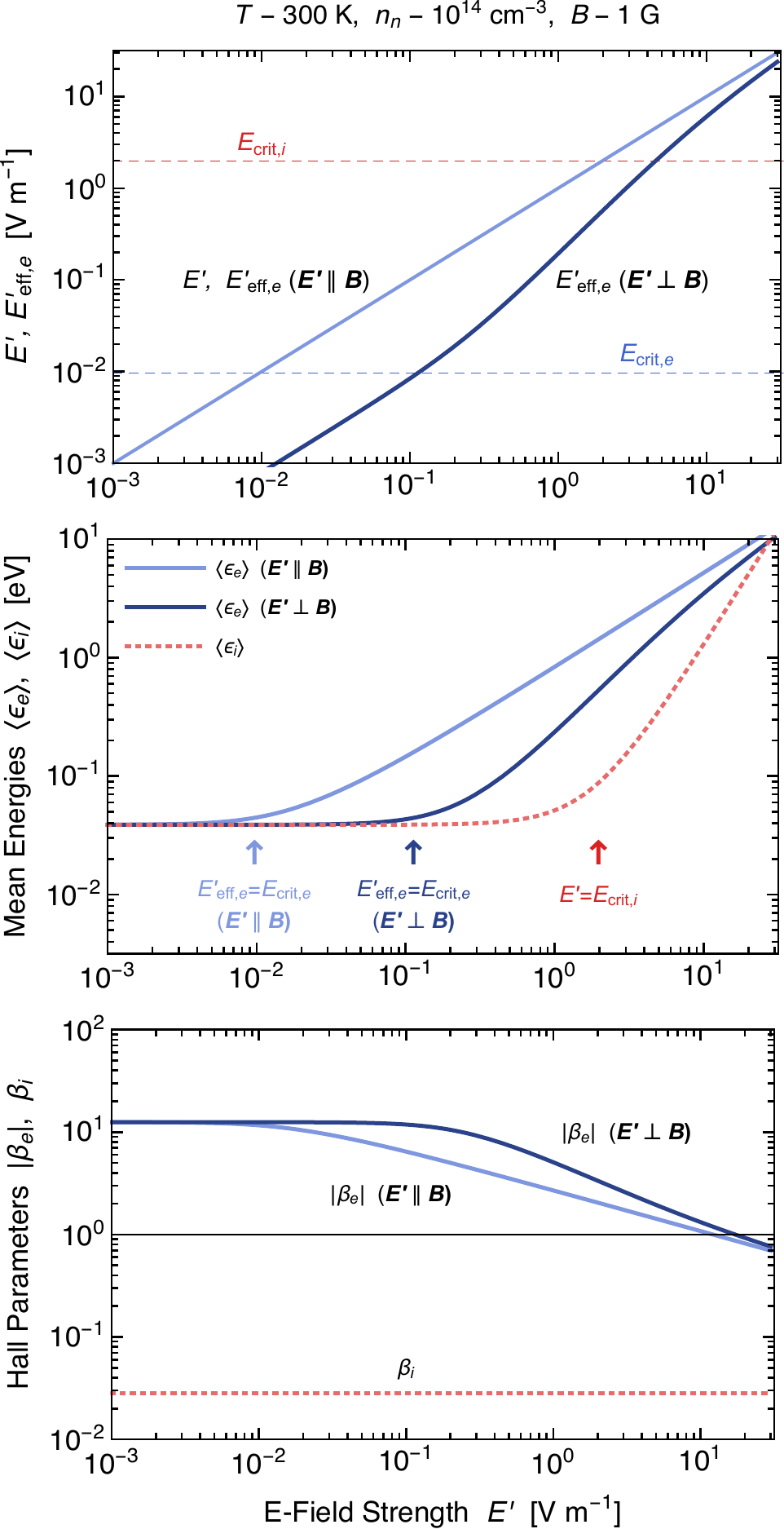}}
\caption{Effective electric field strengths, mean energies, and Hall parameters
as a function of the electric field strength $E'$ for $T=300~{\rm K}$, $n_n=10^{14}~{\rm cm^{-3}}$, and $B=1~{\rm G}$. 
Top panel: $E'_{\rm {eff},e}$ for ${\bm E}' \parallel {\bm B}$ (solid black line) and for ${\bm E}' \perp {\bm B}$ (solid blue line),
compared with the critical field strength for electron heating, $E'_{\rm {eff},e}$ (dashed blue line). 
Note that $E'_{\rm {eff},e} = E'$ for ${\bm E}' \parallel {\bm B}$, and that 
$E'_{\rm {eff},i} \approx E'$ when $\beta_i \ll 1$ as considered here (see the bottom panel). 
Middle panel: $\brac{\eps_e}$ for ${\bm E}' \parallel {\bm B}$ (dark blue line) and for ${\bm E}' \perp {\bm B}$  (light blue line),
and $\brac{\eps_i}$ (red dashed line).
The light and dark blue arrows indicate $E'_{{\rm eff},e} = E_{{\rm crit},e}$,
and the red arrow $E' = E_{{\rm crit},i}$.
Bottom panel: $|\beta_e|$ for ${\bm E}' \parallel {\bm B}$ (dark blue line) and for ${\bm E}' \perp {\bm B}$  (light blue line),
and $\beta_i$ (red dashed line).
}
\label{fig:eebe}
\end{figure}

\subsection{An Example}\label{sec:fig1}
Figure~\ref{fig:eebe} illustrates how the kinetics of plasma particles 
depends on the relative orientation between ${\bm E}'$ and ${\bm B}$.
Here, we plot the mean energies and Hall parameters of electrons and ions  
as a function of $E'$ for two extreme cases of ${\bm E}' \parallel {\bm B}$ and ${\bm E}' \perp {\bm B}$.
The parameters are chosen to be $T = 300~{\rm K}$, $n_n = 10^{14}~{\rm cm^{-3}}$, 
and $B = 1~{\rm G}$.
The values of $T$ and $n_n$ are close to those of the optically thin minimum-mass solar nebula model of \citet{H81} 
as measured at Earth's orbit.
The value of $B$ has been chosen so that the plasma beta $\beta_{\rm plasma} \equiv 8\pi m_n n_n c_s^2/B^2$, 
the ratio of the gas pressure to the magnetic pressure, is set to be $\approx 100$.
For these parameters, one has $\beta_i \approx 0.03 \ll 1$ (see the bottom panel of Figure~\ref{fig:eebe}), 
and hence $\brac{\eps_i}$ is independent of the orientation of ${\bm E}'$.
In contrast, the electron mean energy does depend on the orientation of ${\bm E}$
because $|\beta_e | \approx 1\textrm{--}10$ as long as $E' \la 10~{\rm V~m^{-1}}$. 
In the particular case of ${\bm E}' \perp {\bm B}$, 
the approximation $E'_{{\rm eff},e} \approx E'/|\beta_e|$ holds (see the discussion below Equation~\eqref{eq:Eeff}), 
and therefore heating the electron requires $E' > E_{{\rm crit},e}$ ($\approx 10 E_{{\rm crit},e}$ in the example shown here).

Since $|\beta_e|  \propto \nu_{en}^{-1} \propto \brac{\eps_e}^{-1/2}$,
$|\beta_e|$ decreases with increasing $\brac{\eps_e}$ as shown in the bottom panel of Figure~\ref{fig:eebe}.
This implies that the effects of magnetic fields on the electron conductivity becomes weaker 
as the electric fields heat electrons.
We will come back to this point in Section~\ref{sec:case1}.

\section{Currents in a Strong Electric Field: Analytic Estimates}\label{sec:J}
Assuming that the relaxation timescales of plasma motions and charge reactions are short compared to
the dynamical timescale of the neutral gas, the electric current flowing in the gas is approximately 
determined by the electric field in the neutral-comoving frame, the relation known as Ohm's law.
Ohm's law specifies how strong electric field is needed to sustain an electric current of a given strength. 
%
In \citetalias{OI15}, we showed that in dusty environments like protoplanetary disks, 
there are upper limits on the electric current density that can be realized without electrical breakdown.
However, the derivation was limited to the case where the effect of magnetic field on the kinetic of
plasmas is negligible, i.e., $|\beta_\alpha| \ll 1$.
The aim of this section is to show that the same upper limits apply to the magnitude of the electric current 
density even in the presence of magnetic fields.

\subsection{Generalized Ohm's Law}\label{sec:ohm}
For given $\brac{{\bm v}_\alpha}$ and $n_\alpha$, the electric current density can be written as 
\beq
{\bm J} = \sum_\alpha {\bm J}_\alpha, \qquad {\bm J}_\alpha = q_\alpha n_\alpha \brac{{\bm v}_{\alpha}}.
\label{eq:J_def}
\eeq
When a magnetic field is present,  $\brac{{\bm v}_\alpha}$ and hence ${\bm J}_\alpha$ are
no longer parallel to ${\bm E}'$ because of the magnetic Lorentz force \citep[e.g.,][]{NU86,WN99}.
It follows from Equation~\eqref{eq:valpha} that ${\bm J}_\alpha$ has a general form, 
often called the generalized Ohm's law,
\beq
{\bm J}_\alpha = \sigma_{O,\alpha} {\bm E}'_\parallel  + \sigma_{H,\alpha} \hat{\bm B} \times {\bm E}'_\perp 
+ \sigma_{P,\alpha} {\bm E}'_\perp  ,
\label{eq:generalohm_alpha}
\eeq
where the coefficients 
\beq
\sigma_{O,\alpha} = \frac{q_\alpha^2 n_\alpha}{\mu_{\alpha n}{\nu_{\alpha n}}},
~~
\sigma_{H,\alpha} = - \frac{\beta_\alpha \sigma_{O,\alpha}}{1+\beta_\alpha^2} ,
~~ 
\sigma_{P,\alpha} = \frac{\sigma_{O,\alpha}}{1+\beta_\alpha^2}
\label{eq:sigma}
\eeq
represent the contributions of charge species $\alpha$ to the Ohmic, Hall, and Pedersen
conductivities, respectively.
The generalized Ohm's law reduces to the standard Ohm's law ${\bm J}_\alpha = \sigma_{O,\alpha}{\bm E}'$ 
in the limit of $|\beta_\alpha| \to 0$. 

The expression of the generalized Ohm's law in the vector form is much more complex than 
that of the standard Ohm's law.
However, one can show from Equation~\eqref{eq:generalohm_alpha} that 
the {\it magnitude} of the current, $J_\alpha \equiv |{\bm J}_\alpha|$, has a much simpler expression
\beq
J_\alpha = \sigma_{O,\alpha} E_{{\rm eff},\alpha}',
\label{eq:Jabs}
\eeq
where $E_{{\rm eff},\alpha}'$ is the effective electric field strength already introduced in Equation~\eqref{eq:Eeff}.
Equation~\eqref{eq:Jabs} is formally identical to the scalar version of the standard Ohm's law, 
$J_\alpha = \sigma_{O,\alpha} E'$, 
except that $E'$ has now been replaced by $E_{{\rm eff},\alpha}'$.
As discussed in Section~\ref{sec:kinetics}, $\nu_{\alpha n}$ for ions is constant, 
and that for electrons depends on ${\bm B}$ and ${\bm E}'$ only through $E_{{\rm eff},e}'$.
Therefore, if $n_\alpha$ depends only on $E_{{\rm eff},\alpha}'$, so does $J_\alpha$, and 
the dependence is identical to that of $J_\alpha$ on $E'$ for vanishing magnetic field.

\subsection{Upper limits on the Currents in a Dusty Gas}\label{sec:Jmax}
Protoplanetary disks are weakly ionized plasmas with an typical ionization fraction 
much below $10^{-10}$ in their inner part.
They are also dusty plasmas where small dust grains affect the plasma densities
and can even contribute to the overall charge neutrality of the dust--gas mixture 
\citep[e.g.,][]{SMUN00,IN06a,W07,BG09,O09}.
As highlighted in \citetalias{OI15}, plasma heating accelerates the adsorption of the plasma particles 
onto small dust grains, thus suppressing the electric currents in the mixture.
We here show that the same happens even in the presence of magnetic fields.

When small dust grains are so abundant that electron capture by the grains outpaces 
 electron--ion recombination in the gas, $n_e$ is approximately given by  
\citepalias[Equation~(58) of][]{OI15}
\beq
n_e \approx \frac{\zeta n_n}{\pi a^2 n_d \brac{v_e} C_e},
\label{eq:ne_equil}
\eeq
where $\zeta$ is the ionization rate of the neutral gas, 
 $a$ and $n_d$ are the size and number density of the grains,
and $C_e$ is a dimensionless factor typically in the range $0.01$--1.  
In Equation~\eqref{eq:ne_equil}, the factor $\pi a^2 n_d \brac{v_e}$ corresponds to 
the rate of electron capture for neutral grains, 
while $C_e$ expresses how much the electron--grains collisions are suppressed 
when the grains are negatively charged (see Equation~\eqref{eq:Ce} 
for its expression under the Maxwellian approximation).
Since $\brac{v_e} \propto \brac{\eps_e}^{1/2}$ (see Equation~\eqref{eq:ve}),
$n_e$ decreases with increasing $E_{{\rm eff},e}'$ at $E_{{\rm eff},e}' > E_{{\rm crit},e}$.

When Equation~\eqref{eq:ne_equil} holds,
the magnitude of the electron current has a useful limiting expression in the strong field limit $E' \gg E_{{\rm crit},e}$.
Substitution of Equation~\eqref{eq:ne_equil} together with Equations~\eqref{eq:ve} and \eqref{eq:nu_en}
into $J_e = \sigma_{O,e}E'_{{\rm eff},e}$ gives 
\beq
J_e \approx  \frac{\zeta  e^2}{\pi a^2n_d C_e}
\frac{9\pi E_{{\rm eff},e}'}{64  \sigma_{en} \brac{\eps_e}}.
\eeq
For $E_{{\rm eff},e}' \gg E_{{\rm crit},e}$, 
Equation~\eqref{eq:ee_general} has a limiting expression
\beq
\brac{\eps_e} \approx \frac{3}{16}\sqrt{\frac{3\pi P_\ell m_n}{2m_e}} \frac{e E_{{\rm eff},e}' }{n_n \sigma_{en}},
\eeq
and hence we obtain $J_e \approx J_{e,{\rm max}}$, where 
\beqn
 J_{e,{\rm max}} \equiv  \frac{1}{C_e}\sqrt{\frac{3\pi m_e}{8P_\ell m_n}} \frac{\zeta e n_n}{\pi a^2 n_d}.
\label{eq:Jemax}
\eeqn
Equation~\eqref{eq:Jemax} is almost identical to Equation (60) of \citetalias{OI15},
but the new expression accounts for the inelastic energy losses in electron--neutral collisions. 

The most important property of $J_{e,{\rm max}}$ is that it does not depend on the electric field strength 
except through $C_e$ and $P_\ell$. 
Ignoring the dependence of $C_e$ and $P_\ell$ on $E_{{\rm eff},e}$, 
the electron conductivity $\sigma_{O,e}$ is {\it inversely} proportional to $E'_{{\rm eff},e}$,
because both $n_e$ and $\nu_{en}^{-1}$ scale as $\brac{v_e}^{-1} \propto (E'_{{\rm eff},e})^{-1/2}$.
This cancels the linear dependence of $J_e = \sigma_{O,e}E'_{{\rm eff},e}$ on $E'_{{\rm eff},e}$.
In fact, $J_{e,{\rm max}}$ increases or decreases slowly with $E'_{{\rm eff},e}$ depending on the behavior 
of $P_\ell$ and $C_e$ as we demonstrate in Section~\ref{sec:case1}.

Similarly, the limiting value of the ion current $J_i$ at $E'_{{\rm eff},i} \gg E_{{\rm crit},i}$ 
is given by \citep{JO18}
\beq
J_{i,{\rm max}} = \frac{1}{C_i} \frac{\zeta e n_n}{\pi a^2 n_d} ,
\label{eq:Jimax}
\eeq
where $C_i$ expresses the enhancement of the ion--grain collision 
frequency due to their Coulomb attraction
 (see Equations~\eqref{eq:Ci} and \eqref{eq:vi} 
for its expression under the Maxwellian approximation).
As we demonstrate in Section~\ref{sec:case}, 
$J_i$ dominates the total current only when negatively charged small grains are so abundant 
that the electrons in the gas gets depleted compared to the ions in the gas.
In that case, it is usually safe to assume $C_i \approx 1$ 
(\citetalias{OI15}; \citealt{JO18}). 

If we rewrite $n_n/n_d$ in terms of the dust-gas mass ratio $f_{\rm dg} \equiv m_d n_d/(m_n n_n)$, 
where $m_d = 4\pi \rho_{\rm int} a^3/3$ and $\rho_{\rm int}$ are the mass and internal density of the grains, 
respectively, we obtain
\beqn
J_{e,{\rm max}} &\approx& 1.4\times 10^{-8} \pfrac{0.02}{C_e}\pfrac{1}{P_\ell}^{1/2}
\pfrac{10^{-6}}{f_{\rm dg}}\pfrac{\rho_{\rm int}}{3~\rm g~cm^{-3}}\pfrac{a}{0.1~\micron}
\nonumber \\
&& \times \pfrac{\zeta}{10^{-18}~\rm s^{-1}}  ~\rm A~m^{-2} ,
\eeqn
\beqn
J_{i,{\rm max}} &\approx& 1.6\times 10^{-8} \pfrac{1}{C_i}
\pfrac{10^{-6}}{f_{\rm dg}}\pfrac{\rho_{\rm int}}{3~\rm g~cm^{-3}}\pfrac{a}{0.1~\micron}
\nonumber \\
&& \times \pfrac{\zeta}{10^{-18}~\rm s^{-1}}  ~\rm A~m^{-2} .
\label{eq:Jimax}
\eeqn

\section{Currents in a Strong Electric Field: Calculations with an Ionization Model}\label{sec:case}
We here demonstrate that the magnitude of the current density in a dusty gas 
indeed plateaus out at high electric field strengths.
To do that, we compute $n_\alpha$ as a function of the electric field strength 
consistently with the velocity distribution of plasma particles.
We also study how the electric conductivities and magnetic resistivities (to be introduced in Section~\ref{sec:eta})
depend on the electric field strength. 

\subsection{Ionization Model}\label{sec:ionization}
We employ a simplified charge reaction model developed in \citetalias{OI15}. 
In this model, we only consider one species of positive ions and one species of charged dust grains. 
The grains are assume to have a narrow charge distribution peaked at the mean charge $Ze$, 
and the dispersion of the grain charge is neglected.
The charge reactions we consider are the ionization by external ionizing sources,
recombination of the electrons and ions in the gas, and adsorption of the plasma particles to the grains. 

Under these assumptions, the rate equations for the ion and electron number densities $n_i$ and $n_e$ 
are given by
\beq
\frac{dn_i}{dt} = \zeta n_n - K_{\rm rec} n_i n_e - K_{di}n_dn_i,  
\label{eq:dnidt}
\eeq
\beq
\frac{dn_e}{dt} = \zeta n_n  - K_{\rm rec} n_i n_e - K_{de}n_dn_e,  
\label{eq:dnedt}
\eeq
respectively, where $\zeta$ is the rate of external ionization,  
and $K_{d\alpha}~(\alpha=i,e)$ and $K_{\rm rec}$ are
the rate coefficients for gas-phase recombination and plasma adsorption onto the grains, respectively.
The dimensionless numbers $C_e$ and $C_i$ appearing in Equations~\eqref{eq:Jemax}
and \eqref{eq:Jimax} are related to $K_{de}$ and $K_{di}$ 
as $C_e = K_{de}/(\pi a^2 \brac{v_e})$ and $ C_i = K_{di} /(\pi a^2 \brac{v_i})$, 
respectively, where $\brac{v_i}$ is the mean speed of the ions (see Appendix~\ref{sec:K}).
The reaction rate coefficients depend on the velocity distribution of ions and electrons, 
and we use the expressions under the offset Maxwell approximation given in Appendix~\ref{sec:K}.
The adsorption rate coefficients depend on the grain charge $Z$, which is 
related to $n_i$ and $n_e$ via the charge neutrality of the gas--dust mixture, 
\beq
n_i - n_e + Z n_d = 0.
\label{eq:neut}
\eeq

We assume steady state $dn_i/dt = dn_e/dt = 0$ and 
solve Equations~\eqref{eq:dnidt}--\eqref{eq:neut} for $n_i$, $n_e$, and $Z$ 
under the steady state conditions as a function of $E'$ and the relative orientation 
between ${\bm E'}$ and ${\bm B}$ (below assumed to be either parallel or perpendicular to each other).
The solution of the equations is searched for using the semianalytic approach detailed in Section 3.2.4 of \citetalias{OI15}.
The solution generally satisfies $Z < 0$.  

The adopted model neglects the ionization of the neutral gas by electrically heated electrons. 
In \citetalias{OI15}, we showed that this impact ionization dominates over external ionization  
when $\brac{\eps_e} \ga 3~\rm eV$, for which the number of electrons with a kinetic energy 
above the ${\rm H_ 2}$ ionization potential of $15.4~\rm eV$ is substantial. 
When the electric field is strong enough to fulfill this condition, 
the impact ionization causes electrical breakdown, leading to an abrupt increase in the electric current similar to lightning.
However, we found in \citetalias{OI15} that this lightning-like current is unstable to perturbations 
when the charged grains are the dominant negative charge carriers. 
Because the stability issue of the lightning-like discharge is not the focus of this paper, we simply neglect impact ionization 
and instead restrict the electric-field strength to below the threshold corresponding to $\brac{\eps_e} = 3~\rm eV$. {Note that it is safe to use our approximate expression for $P_\ell$ (Equation~\eqref{eq:Pell_model}) below this threshold.}

\subsection{Parameter Choice} \label{sec:param}
As in Section~\ref{sec:fig1}, we consider the inner part of protoplanetary disks 
and adopt $T = 300~\rm K$ and $n_n = 10^{14}~{\rm cm^{-3}}$. 
The ionization rate $\zeta$ is taken to be $\zeta = 10^{-18}~{\rm s^{-1}}$
assuming that the gas is mainly ionized by short-lived radionuclides 
(for which $\zeta\sim 10^{-20}$--$10^{-18}~\rm s^{-1}$; see \citealt{UN81,UN09}; \citealt{S92}; \citealt{CABV13}). 
This is a reasonable assumption for the dense part of disks where external ionizing {sourses} such as 
cosmic rays and X-rays are greatly attenuated.
Dust grains are {assumed} to have a radius of $a = 0.1~\micron$. 

We consider two cases of $f_{\rm dg} = 10^{-6}$ and $10^{-4}$ (henceforth cases 1 and 2, respectively).
As discussed by \citet{O09} and in \citetalias{OI15}, the ionization state of a gas--dust mixture
depends on whether  $n_e > |Z|n_d$ or $n_e < |Z|n_d$, i.e., whether 
the dominant negative charge carriers are the electrons in the gas or the negatively charged grains.
We have selected the two values of $f_{\rm dg}$ so that 
the two extreme conditions  $n_e \gg |Z|n_d$ and $n_e \ll |Z|n_d$ are realized in cases 1 and 2, respectively,
in the limit of $E' \to 0$.
If $n_e \gg |Z|n_d$, Equation~\eqref{eq:neut} gives $n_i \approx n_e$, 
and one generally has $J_e \gg J_i$ because electrons are much more mobile than ions.
In the opposite case of $|Z|n_d \gg n_e$, the number of electrons in the gas 
is typically two orders of magnitude smaller than that of ions in the gas,   
and for this reason $J_i$ can be comparable to or even dominate over $J_e$. 
The Coulomb reduction factor $C_e$ is determined by which regime applies, 
with $C_e \approx 0.02$--0.1 for $n_e \gg |Z|n_d$ and $C_e \approx 1$ for $n_e \ll |Z|n_d$
(see Figure 9 of \citetalias{OI15}).

\subsection{Magnetic Resistivities}\label{sec:eta}
To quantify the non-ideal MHD effects, it is useful to 
introduce the magnetic resistivities for Ohmic diffusion, Hall drift, and ambipolar diffusion 
defined by (see, e.g., \citealt{W07})
\beq
\eta_O = \frac{c^2}{4\pi\sigma_O},
\label{eq:etaO}
\eeq
\beq
\eta_H = \frac{c^2 \sigma_H}{4\pi(\sigma_H^2+\sigma_P^2)},
\label{eq:etaH}
\eeq
\beq
\eta_A = \frac{c^2 \sigma_P}{4\pi(\sigma_H^2+\sigma_P^2)} - \eta_O,
\label{eq:etaA}
\eeq
respectively, where $\sigma_O = \sum_\alpha \sigma_{O,\alpha}$, 
$\sigma_H = \sum_\alpha \sigma_{H,\alpha}$, and $\sigma_P = \sum_\alpha \sigma_{P,\alpha}$
are the Ohmic, Hall, and Pedersen conductivities accounting for the contributions of all charged species.
In general, the dominant non-ideal effect has the largest resistivity. 

If $n_i \approx n_e$ as in case 1, the resistivities satisfy simple relations \citep{W07}
\beq
\eta_O \approx \frac{c^2}{4\pi\sigma_{O,e}}, \quad \eta_H \approx |\beta_e|\eta_O, 
\quad \eta_A \approx \beta_i\eta_H \approx \beta_i|\beta_e|\eta_O,
\label{eq:eta_simple}
\eeq
and therefore all the resistivities simply scale as $\sigma_{O,e}^{-1}$. 
Furthermore, if $\beta_i < 1 < |\beta_e|$, then $\eta_H$ is larger than $\eta_O$ and $\eta_A$, 
indicating that Hall drift dominates over Ohmic and ambipolar diffusion. 
If $n_i \gg n_e$ as in case 2, the dependences of $\eta_H$ and $\eta_A$ are generally complex \citep[see, e.g.,][]{XB16}.
In contrast, $\eta_O$ is always inversely proportional to $\sigma_{O}$, 
and hence to the number densities of charged particles, no matter whether $n_i = n_e$ or not. 

\subsection{Results: Case 1} \label{sec:case1}
\begin{figure*}
\centering
\resizebox{\hsize}{!}{\includegraphics{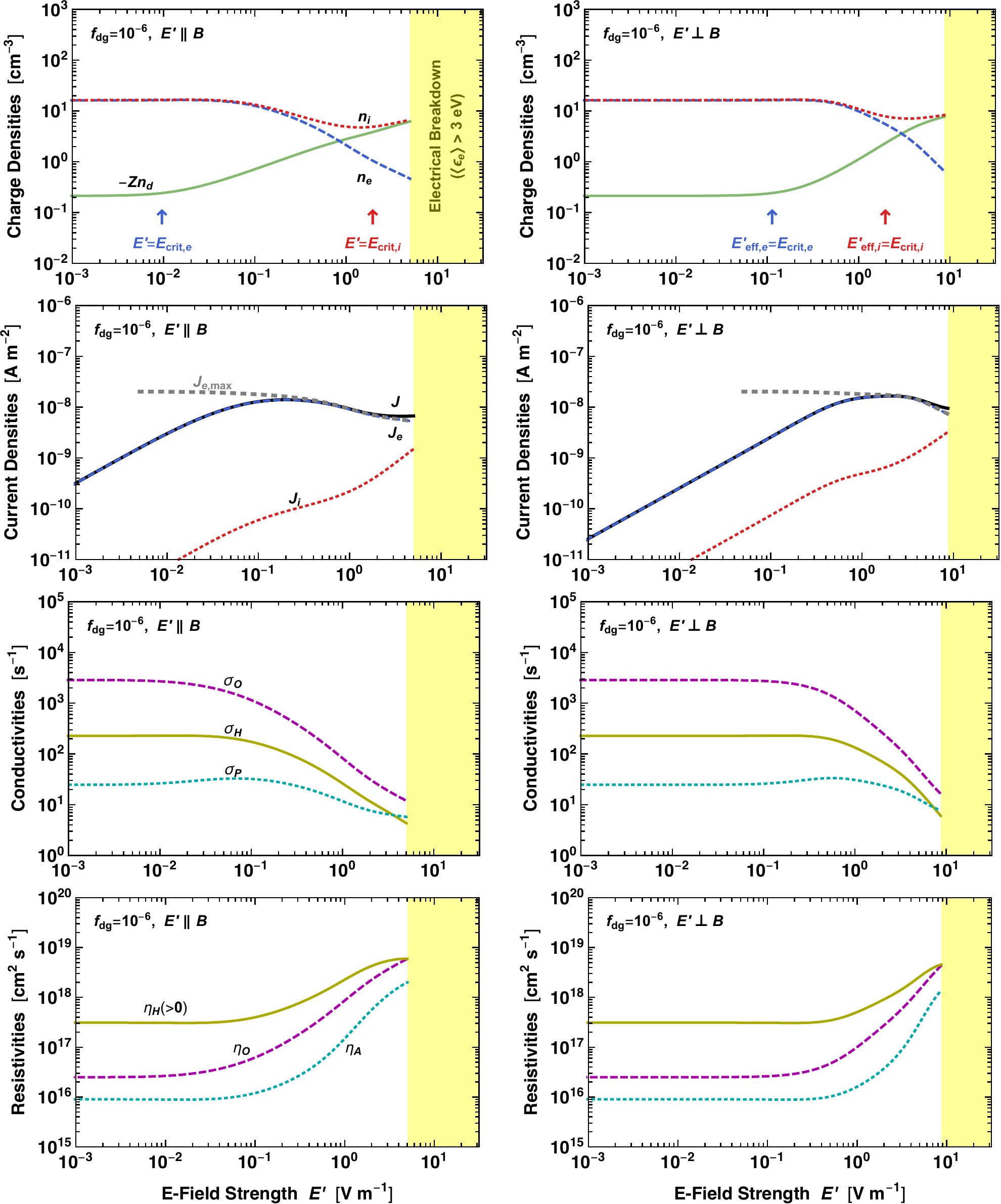}}
\caption{Number densities of charges in the gas phase and on dust grains (top row), 
electric currents (second row), electric conductivities (third row), and magnetic diffusivities (bottom row)
as a function of the electric field strength $E'$ for case 1 ($f_{\rm dg} = 10^{-6}$). 
The left and right columns are for electric field parallel and perpendicular to the magnetic field, respectively.  
{The yellow shaded area marks the field range where $\brac{\eps_e} > 3~\rm eV$, at} which electrical breakdown of the gas 
occurs due to impact ionization by hot electrons (see \citetalias{OI15}). 
The blue and red arrows in the top row indicate $E'_{{\rm eff},e} = E_{{\rm crit},e}$
and $E'_{{\rm eff},i} = E_{{\rm crit},i}$, respectively.
The thick dashed lines in the second row show the maximum electron current $J_{{\rm max},e}$ given by Equation~\eqref{eq:Jemax},
with $C_e$ given by Equation~\eqref{eq:Ce}.
Because  $J \approx J_e$ in these particular examples, the lines for $J$ and $J_e$ in the second row
overlap. 
}
\label{fig:1e-6}
\end{figure*}

We begin by studying how plasma heating changes the charge reaction balance in this case. 
The top row of Figure \ref{fig:1e-6} shows the number densities of ions and electrons in the gas, 
$n_i$ and $n_e$, as well as the number density of electrons adsorbed on grain surfaces, $-Zn_d$,
as a function of $E'$ for two extreme orientations of ${\bm E}'$ relative to ${\bm B}$. 
As mentioned in Section~\ref{sec:param}, case 1 is designed so that the condition $n_i \approx n_e \gg |Z|n_d$ holds in the limit of small $E'$.  
As $E'$ increases and the electric heating of electrons sets in ($E'_{{\rm eff},e} = E_{{\rm crit},e}$), 
$|Z|$ starts to increase because the heated electrons collide with and adsorb onto dust grains more frequently \citepalias{OI15}. 
This also causes the decrease of $n_e$ with increasing $E'$.   
In this particular example, the charged grains become the dominant negative charge carriers $|Z|n_d > n_e$
at $E' \ga 1~\rm V~m^{-1}$.
 
The second row of Figure~\ref{fig:1e-6} shows the magnitude of the total current density, $J$, 
as well as of the ion and electron current densities, $J_i$ and $J_e$. 
As stated in Section~\ref{sec:param}, the electron current dominates in the case of $n_e \approx n_i$. 
The results shown here demonstrate that the total current for such a case approaches $J_{{\rm max},e}$ 
given by Equation~\eqref{eq:Jemax}, irrespective of the orientation of ${\bm E}'$. 
It is important to note that the value of $E'$ required to heat electrons does depend on its orientation, 
with the ${\bm E}' \perp {\bm B}$ case requiring 10 times higher $E'$ than the ${\bm E}' \parallel {\bm B}$ case 
(see the blue arrows in the top panels of Figure~\ref{fig:1e-6} for the onset of electron heating).
This illustrates that $J$ should be viewed as a function of $E'_{{\rm eff},e}$ rather than of $E'$ 
as long as the electric heating of ions is negligible.

The bottom two rows of Figure~\ref{fig:1e-6} plot the conductivities and resistivities as a function of $E'$.
Because $n_i \approx n_e$, the resistivities satisfy the relations 
given by Equation~\eqref{eq:eta_simple}. 
In this particular case, one has $\beta_i < 1 < |\beta_e|$ (see Figure~\ref{fig:eebe}), 
and therefore $\eta_H$  is the largest.
Because $\sigma_{O,e}$ decreases with increasing $E'$ \footnote{This 
follows from $\sigma_{O,e} \propto (C_e P_\ell^{1/2} E'_{{\rm eff},e})^{-1}$, 
$P_\ell \propto \brac{\eps_e}^{-1/2} \propto (E'_{{\rm eff},e})^{-1/2}$, 
and $C_e$ being an increasing function of $E'_{{\rm eff},e}$.
}, all resistivities increase with $E'$. 
Moreover, since $|\beta_e|$ is a decreasing function of $E'$ (Figure~\ref{fig:eebe}), 
$\eta_H \approx |\beta_e|\eta_O$ increases more slowly than $\eta_O$, 
and hence the difference between the two resistivities decrease toward higher $E'$.
 

\subsection{Results: Case 2}
\begin{figure*}
\centering
\resizebox{\hsize}{!}{\includegraphics{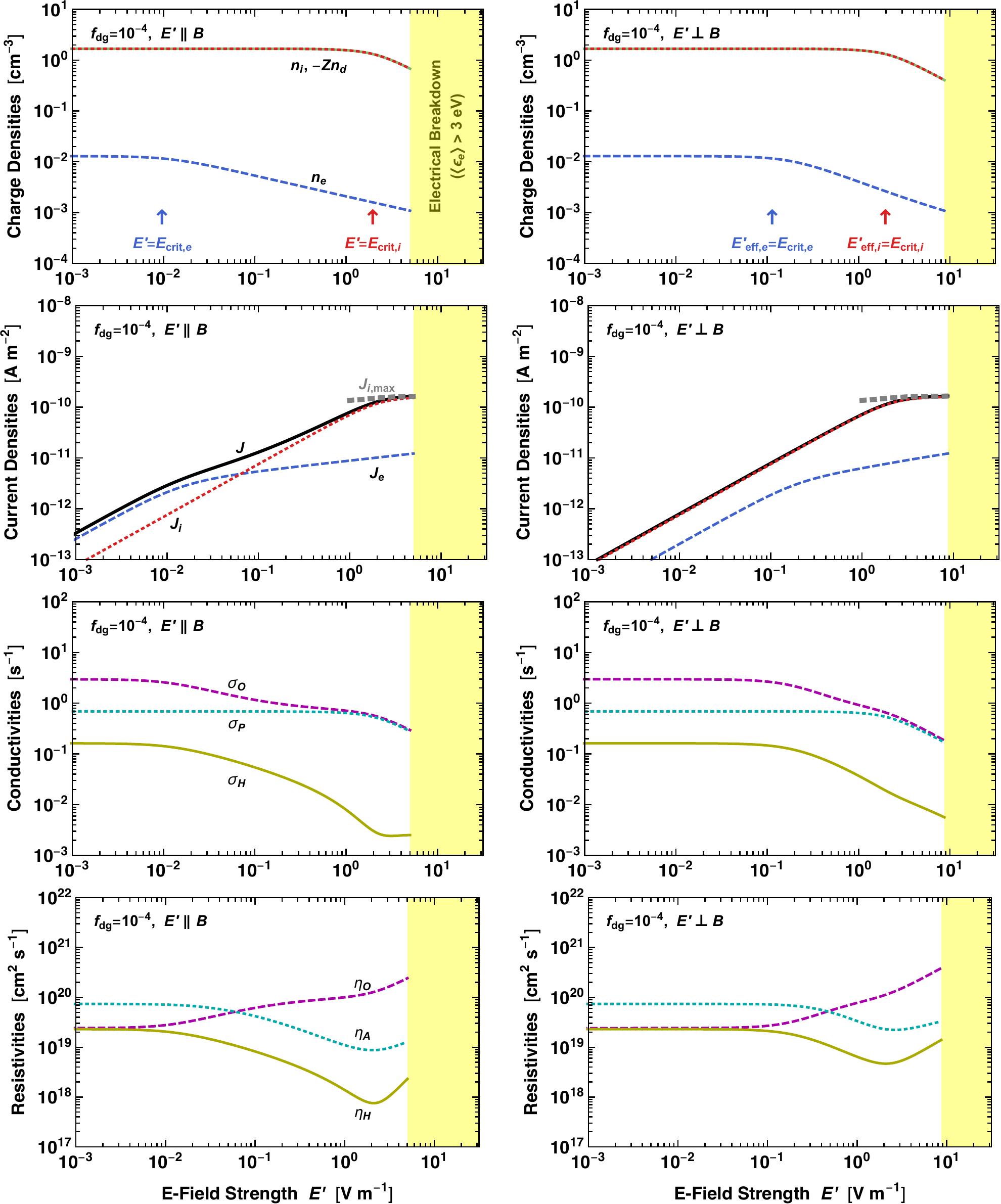}}
\caption{Same as Figure~\ref{fig:1e-6}, but for case 2 ($f_{\rm dg} = 10^{-4}$). 
The thick dashed lines in the second row indicate the maximum ion current $J_{i,{\rm max}}$ given by Equation~\eqref{eq:Jimax},
with $C_i$ given by Equation~\eqref{eq:Ci}.
}
\label{fig:1e-4}
\end{figure*}
Case 2 serves as an example where the condition $n_i \approx |Z|n_d \gg n_e$ holds (top row of Figure~\ref{fig:1e-4}).
At $E'_{{\rm eff},e} > E_{{\rm crit},e}$, $n_e$ decreases with increasing $E'$ for the reason described in Section~\label{sec:case1}.
For the same reason, $n_i$ also decreases at $E'_{{\rm eff},i} > E_{{\rm crit},i}$.

Because $n_i \gg n_i$ in case 2, $J_i$ gives a relatively large contribution 
to $J$ as shown in the second row of Figure~\ref{fig:1e-4}.
For ${\bm E}' \parallel {\bm B}$, $J_i$ dominates over $J_e$ as the latter plateaus out at $J_{e,{\rm max}}$.
At $E' > E_{{\rm crit},i}$, $n_i$ also decreases with $E'$, and consequently $J_i$ relaxes into a constant $J_{i,{\rm max}}$ 
given by Equation~\eqref{eq:Jimax} as predicted in Section~\ref{sec:Jmax}.

The bottom two rows of Figure~\ref{fig:1e-4} show the conductivities and resistivities versus $E'$ for case 2. 
As in case 1, the conductivities decrease monotonically with increasing $E'$,
and hence $\eta_O$ $(\propto \sigma_O^{-1})$ increases with $E'$.
In contrast, $\eta_A$ and $\eta_H$ {\it decrease} until $E'$ reaches $E_{{\rm crit},i}$, contrary to case 1.
As a consequence, $\eta_O$ dominates over $\eta_A$ at $E' \ga 10E_{{\rm crit},e}$.

The reason why $\eta_A$ and $\eta_H$ decrease with $E'$ is the following.  
In case 2, electrons are so depleted from the gas phase that  $\sigma_P \approx \sigma_{P,i}$.
In contrast, $\sigma_H$ is still dominated by $\sigma_{H,e}$, and is smaller than $\sigma_{P}$
as shown in the third row of  Figure~\ref{fig:1e-4}.
From these relations, we obtain $\eta_H \propto \sigma_{H,e}/\sigma_{P,i}^2$ 
and  $\eta_A \propto 1/\sigma_{P,i}-1/\sigma_{O}$.
At $E' < E_{{\rm crit},i}$,  $\sigma_{P,i}$ $(\propto n_i)$ is constant,
and therefore $\eta_H \propto \sigma_{H,e}$ is a decreasing function of $E'$. 
In addition, at $E' \ga 10E_{{\rm crit},e}$, the electrons in the gas 
are further depleted so that $\sigma_O$ approaches $\sigma_{O,i}$.
However, because $\beta_i \ll 1$ in this example (see Equation~\eqref{eq:sigma}), 
$\sigma_{O,i}$ is approximately equal to $\approx \sigma_{P,i}$. 
Therefore, $\eta_A \propto 1/\sigma_{P,i}-1/\sigma_{O}$ vanishes 
as $\sigma_{O}$ approaches $\sigma_{O,i}$.
 
As is obvious from the above analytic argument, how $\eta_H$ and $\eta_A$ behave 
as a function of $E'$ depends on the values of $\beta_i$ and $\beta_e$ as well as on $n_e/n_i$. 
Exploring these dependences over a wide parameter space will be interesting future work, but is not the subject of this paper. 
It is perhaps more important to emphasize that the Ohmic resistivity $\eta_O$ never decreases with $E'$,
and that it tends to dominate over the other two resistivities at sufficiently large $E'$. 
The second point is a natural consequence of the general property that Ohmic diffusion dominates when $|\beta_e|$ falls below $1$
\citep{XB16}. Electron heating causes a decrease in $|\beta_e|$, and hence tends to make Ohmic diffusion the dominant {nonideal} MHD effect. 

\section{Implications for MHD in Protoplanetary Disks}\label{sec:discussion}
The most important finding of this study is that 
plasma heating places upper limits on the electric currents 
even when magnetic fields strongly affect the plasma motions, 
corresponding to the case where Hall drift or ambipolar diffusion dominates over Ohmic diffusion. 
Any MHD motion that can produce a current larger the limits 
would either cause electrical breakdown of the gas \citep{IS05,MOI12}
or would get suppressed before the breakdown sets in \citep{MMOI17}. 
The upper limits decrease with increasing the abundance of small dust grains. 
Moreover, it is shown by \citepalias{OI15} that if the grains are the dominant negative charge carriers, 
the breakdown current is unstable to perturbations. 
Therefore, we can speculate that any MHD motion that produces a current exceeding the limits  
would be stably sustained only if small  grains are heavily depleted.

The next question is then whether the MHD motions of real protoplanetary disks 
would indeed produce such a high electric current.  
In \citetalias{OI15} and \citet{MO16}, we pointed out that the small-scale currents produced 
by  fully developed MRI turbulence can indeed exceed the limits. 
However, as mentioned in Section~\ref{sec:intro}, recent studies show that 
MRI-driven turbulence is unlikely to operate in most part of the disks 
if all three non-ideal MHD effects are taken into account. 
Based on the current understanding of the MHD in protoplanetary disks, 
we here consider more coherent gas motions on a larger scale.
A candidate that drives such a motion is the HSI, 
which has recently been found to generate a large-scale magnetic field \citep{KL13,B14,B17,LKF14}.
The MHD simulations by \citet{B15,B17} 
show that a large-scale field produced by the HSI is in some cases 
accompanied by a strong current layer near the midplane 
(see Section 5.1 of \citealt{B15}; Section 5.2 of \citealt{B17}).
Because the strong magnetic fields generated by the HSI can provide 
a high level of accretion stress, it is important to assess whether the upper limits 
on the electric current could affect the saturation level of the HSI.
We note that a strong current can also occur on the disk surface where the ionization rate 
is higher than in the midplane \citep[e.g.,][]{BS13b,GTNM15}, 
but such a current is less likely to be relevant because the upper limits increase with ionization rate
(see Equation~\eqref{eq:Jemax} and \eqref{eq:Jimax}).

To estimate the magnitude of the current associated with such a large-scale gas motion, 
we make use of Ampere's law 
\beq
{\bm J} = \frac{c}{4\pi}\nabla \times {\bm B}.
\label{eq:Ampere}
\eeq
We assume that the magnetic field produce by the HSI has typical magnitude $\bar{B}$ 
and length scale $L$ over which the magnetic field lines are bent. 
Using Equation~\eqref{eq:Ampere}, one can estimate that 
the electric current that produces the magnetic field has a typical current density of 
$\bar{J} \sim {c\bar{B}}/(4\pi L)$. 
We rewrite this as 
\beqn
\bar{J} &\sim& \sqrt{\frac{\rho}{2\pi \beta_{\rm pl}}} \frac{c\Omega H}{L}
\nonumber \\
&\sim& 10^{-8}\pfrac{H}{L} \pfrac{\rho}{10^{-9}~\rm g~cm^{-3}}^{1/2}\pfrac{100}{\beta_{\rm pl}}^{1/2}
\pfrac{\Omega}{2\pi ~\rm yr^{-1}} ~\rm A~m^{-2}, \qquad
\label{eq:J_HSI}
\eeqn
where $\beta_{\rm pl} = 8\pi \rho c_s^2/\bar{B}^2$ 
is the plasma beta and $H = c_s /\Omega$ is the gas scale height, 
with $\rho$, $c_s$, and $\Omega$ being the mass density, 
sound speed, and local orbital frequency of the disk gas, respectively. 
In the case of the HSI, 
the induced magnetic field near the midplane has $\beta_{\rm pl} \sim 100$ and $L\sim H$
\cite[see Figure 8 of][]{B17}.

Now we compare $\bar{J}$ given by Equation~\eqref{eq:J_HSI} with 
the upper limits $J_{e,{\rm max}}$ and $J_{i,{\rm max}}$
given by Equations~\eqref{eq:Jemax} and \eqref{eq:Jimax}. 
As an example, we consider the minimum-mass solar nebula model \citep{W77b,H81}, 
in which $\rho \sim 10^{-9}(r/1~\rm au)^{-11/4}~{\rm g~cm^{-3}}$ at the midplane 
and $\Omega = 2\pi (r/1~\rm au)^{-3/2}~\rm yr^{-1}$, 
with $r$ being the distance from the central star (assumed to be of one solar mass). 
As in Section~\ref{sec:case}, we take $\zeta \sim 10^{-18}~\rm s^{-1}$ and $a_0 \sim 0.1~\micron$, 
and parametrize the amount of small dust grains that contributes charge reactions with 
the dust-to-gas mass ration $f_{\rm dg}$. 
For $\beta_{\rm pl} \sim 100$ and $L\sim H$,
one finds  that $\bar{J}$ exceeds the limits at $r \la 1$ au and 3 au if 
$f_{\rm dg} \sim 10^{-6}$ and $10^{-4}$, respectively.

To summarize, we have found that the upper limits on the electric current imposed by plasma heating
can indeed affect large-scale MHD motions in the inner $\sim 1~\rm au$ of protoplanetary disks.
This could potentially have a substantial influence on the accretion of the inner disk regions.
If such a motion can recurrently trigger electrical breakdown near the midplane, 
this could also potentially lead to 
the formation of chondrules through flash heating in the lightning current \citep[e.g.,][]{W66,DC00}. 
{However, it is also possible that the enhanced Ohmic resistivity at $E' > E_{{\rm crit},e}$ diffuses the midplane current layer before the electric field strength reaches the breakdown threshold.
If this is the case, the upper limits on the electric current would rather act to push the midplane current layer toward one side of the disk surface, as often observed in MHD simulations that assume a high magnetic diffusivity near the midplane \citep[e.g.,][]{BS13b,LKF14,BLF17,MBO19}. 
Self-consistent simulations including both nonideal MHD and plasma heating   by electric fields are needed to investigate whether recurrent discharge or the escape of the current layer from the disk midplane is a more realistic outcome.
}

Even if the electric field induced by gas motions is not strong enough to directly cause electrical breakdown, 
the limits on the currents can help breakdown via charge separation by 
positron emission from dust particles \citep{JO18} and other potential mechanisms.
{This possibility should also be examined} in future MHD simulations including all non-ideal MHD effects and plasma heating. 
Finally, we note that the HSI is absent when the disk's vertical magnetic field is anti-aligned with the rotation axis.
In such a disk, Hall drift acts to weaken the magnetic field, and therefore the effects of plasma heating on disk accretion 
would be less important.

\section{Summary}\label{sec:summary}

Following \citetalias{OI15}, we have studied how plasma heating by strong electric fields
affects the electric current in dusty protoplanetary disks.  
The new formulation presented in this paper fully takes into account the effects of 
magnetic fields on the motion of plasma particles, thus allowing us to treat Hall drift 
and ambipolar diffusion in addition to Ohmic diffusion. 
We have also included the energy losses of electrons through inelastic collisions with neutrals,
which were also neglected in our previous work.  
Our key findings are summarized as follows. 

\begin{enumerate}
\item In the presence of both electric and magnetic fields, the temperature
of a charged species (ions or electrons) 
is determined by its effective electric field strength (Equation~\eqref{eq:Eeff}; see also \citealt{GZS80}). 
Substantial heating of a charged species occurs when its effective field strength 
exceeds the critical field strength, which is also species-dependent (Equations~\eqref{eq:Ecriti} and \eqref{eq:Ecrit}). 


\item The upper limits on the ion and electron currents discovered in \citetalias{OI15}
for the spacial case of zero magnetic field hold even with a magnetic field (Equations~\eqref{eq:Jemax} and \eqref{eq:Jimax}). 
Any electric current exceeding the limits is only realized with electrical breakdown of the gas.

\item A large-scale electric current produced by an MHD motion in the inner 
part of protoplanetary disks can exceed the upper limits (Section~\ref{sec:discussion}).
A potential mechanism that can drive such a motion near the midplane is the HSI, which operates 
when the vertical magnetic flux is aligned with the direction of the disk rotation axis.  
Based on our previous MHD simulations \citep{MOI12,MMOI17},  
we predict that {a strong current layer near the midplane} could {either} trigger recurrent electrical breakdown or get suppressed by the increasing ohmic diffusion with increasing electric field strength. 
{The former outcome could lead to chondrule formation \citep[e.g.,][]{W66,DC00,JO18} and the latter outcome could result in migration of a strong current layer to one side of the disk surface \citep[e.g.,][]{BS13b,LKF14,BLF17,MBO19}.} 
{Self-consistent} MHD simulations including all non-ideal effects and plasma heating {are needed to investigate which outcome occurs in realistic conditions}.  

\item 
Inelastic collisions between electrons and neutrals suppress the electron energy,
but only by a factor of $\la 4$ as long as the electrons are not hot enough to cause gas ionization
(Appendix~\ref{sec:Pell}). 
The inelastic energy losses have little effect on the onset of electron heating. 
\end{enumerate}

\acknowledgments
We are grateful to Xuening Bai for providing us with the data of his MHD simulations \citep{B15,B17},
which motivated us to derive an analytic estimate of the current density presented in Section~\ref{sec:discussion}. 
{We also thank the anonymous reviewer for comments that helped improve the manuscript.}
This work was supported by JSPS KAKENHI Grant Numbers JP16H04081, JP16K17661, JP17K18812, JP18H05438, and JP19K03926. 

\appendix 

\section{A Simple Model for $P_\ell$}\label{sec:Pell}
\begin{figure}[t]
\centering
\resizebox{8cm}{!}{\includegraphics{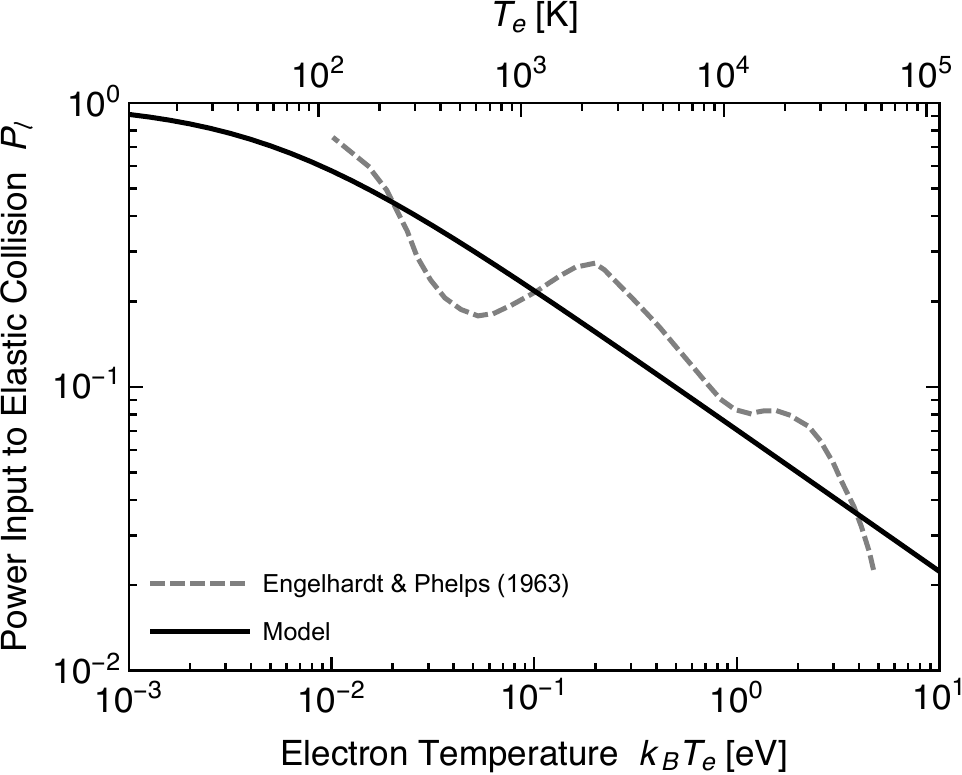}}
\caption{Fractional power input to elastic collision, $P_\ell$, 
as a function of the electron temperature $T_e$. 
The dashed curve shows the theoretical estimate by \citet{EP63} based on experiments (see their Figure 15), 
while the solid curve shows our analytic fit (Equation~\eqref{eq:Pell_model}).
}
\label{fig:Pell}
\end{figure}
We here derive a simple analytic expression for $P_\ell$, the fractional contribution of elastic collisions 
to the electron energy losses through collisions with neutrals, based on the theoretical calculation by \citet{EP63}. 
They derived the fractional power inputs to the elastic and inelastic energy losses 
in ${\rm H}_2$ gas based on electron swarm experiments. 
The inelastic losses include the rotational, vibrational, and electronic excitation losses and the ionization loss.
The dashed curve in Figure~\ref{fig:Pell} shows $P_\ell$ versus the electron temperature $T_e$ taken from Figure 15 
of \citet{EP63}.\footnote{
\citet{EP63} express $P_\ell$ as a function of the ``characteristic energy'' $\eps_K = eD/\mu$,
where $D$ and $\mu$ are the diffusion coefficient and electrical mobility of electrons.
We here assume Maxwellian plasmas, for which $\eps_K$ is equal to $\kB T_e$, known as the Einstein relation.    
}   
Overall, $P_\ell$ decreases from $\approx 1$ to $\sim 0.02$ as 
$T_e$ goes from $10^{-2}~{\rm eV}$ to $10~{\rm eV}$.
The dips in $P_\ell$ at $T_e \sim$ 0.05 eV,  0.1 eV, and 5 eV correspond to 
the energy losses due to the rotational, vibrational, and electronic excitation of ${\rm H_2}$, respectively.
The contribution of the ionization loss is less significant than that of electronic excitation loss as long as 
$T_e \la 10~{\rm eV}$ (see Figure 15 of \citealt{EP63}).
We find that the overall trend can be approximated by a simple function  
\beq
P_\ell = \left(1+\frac{\kB T_e}{0.005~{\rm eV}}\right)^{-1/2},
\label{eq:Pell_model_0}
\eeq
which is shown by the solid curve in Figure~\ref{fig:Pell}. 
Equation~\eqref{eq:Pell_model_0} reproduces the result of \citet{EP63} to within a factor of 2.
{We note that our analytic fit could overestimate $P_\ell$ at $T_e > 4~\rm eV$, where we expect larger energy losses due to electronic excitation and ionization. In fact, Figure 15 of \citealt{EP63} indicates that these energy losses quickly increase at $T_e > 2~\rm eV$.}  
Substitution of $\kB T_e = (2/3)\brac{\eps_e}$ into Equation~\eqref{eq:Pell_model_0} gives
Equation~\eqref{eq:Pell_model} in the main text. 

\begin{figure}[t]
\centering
\resizebox{8cm}{!}{\includegraphics{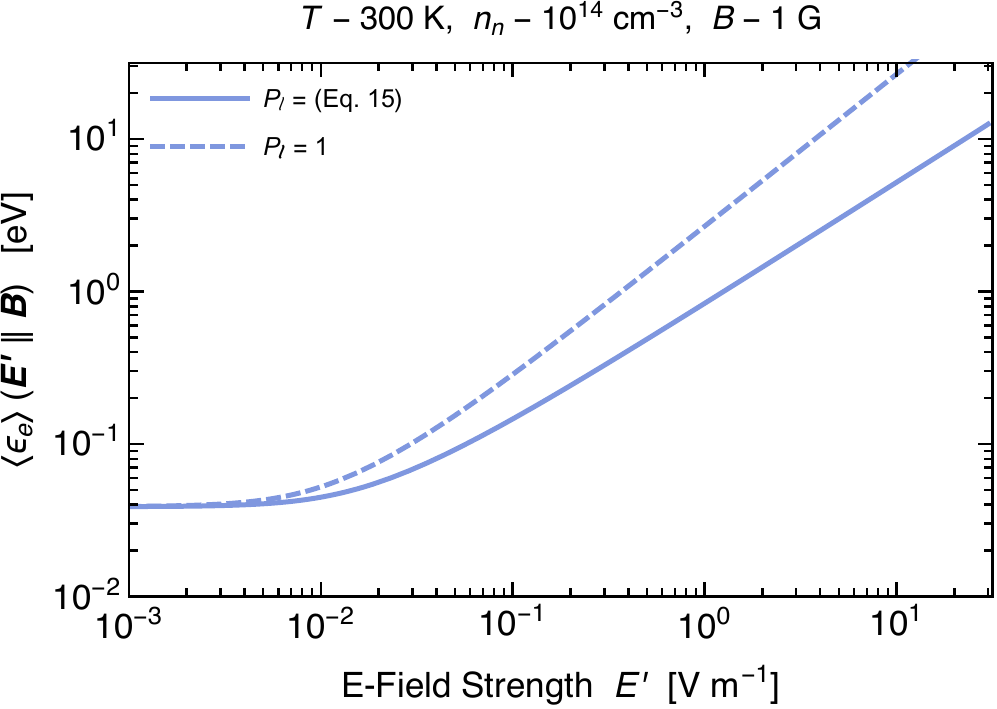}}
\caption{Electron mean energy $\brac{\eps_e}$ versus the electric field strength $E'$ 
in the case of ${\bm E}' \parallel {\bm B}$ for the same parameters as in Figure~\ref{fig:eebe}. 
The solid line assumes $P_\ell$ given by Equation~\eqref{eq:Pell_model}, while  
the dashed line shows how $\brac{\eps_e}$ would behave if all inelastic energy losses were absent, i.e., $P_\ell = 1$.
}
\label{fig:ee_inelastic}
\end{figure}
To illustrate how much the inelastic energy losses affect electron heating, 
we compare in Figure~\ref{fig:ee_inelastic} the mean electron energy $\brac{\eps_e}$ 
(Equation~\eqref{eq:dedt_eff}) for $P_\ell$ given by Equation~\eqref{eq:Pell_model}
with $\brac{\eps_e}$ for purely elastic collisions, $P_\ell = 1$, 
as a function of $E'$.
Here we take $T = 300~{\rm K}$ and $n_n = 10^{14}~{\rm cm^{-3}}$ as in Figure~\ref{fig:eebe}. 
Since $P_\ell \sim 1$ at $T_e \approx T \sim 100~{\rm K}$, 
inelastic losses have little effect on the onset of electron heating.
At $E > \Ecrite$, the electron energy scales as $\brac{\eps_e} \propto E'^{4/5}$, 
which directly follows from the high-field expressions of 
Equations~\eqref{eq:Pell_model} and \eqref{eq:ee_general},  
$P_\ell \propto T_e^{-1/2} \propto \brac{\eps_e}^{-1/2}$
and $\brac{\eps_e} \propto P_\ell^{1/2}E'$.
For example, at $E' = 5~\rm V~m^{-1}$, we obtain $\brac{\eps_e} \approx 3~\rm eV$
with inelastic energy losses, which is about four times lower than $\brac{\eps_e}$ with no inelastic losses. 
Therefore, as long as the electron energy is not high enough to cause electrical breakdown ($\approx 3~\rm eV$), 
the inelastic energy losses only cause a suppression of the electron mean energy by a factor of $\la 4$. 

\section{Reaction Rate Coefficients}\label{sec:K}
The derivation of the adsorption rate coefficients $K_{de}$ and $K_{di}$ 
under the Maxwellian approximation can be found in the literature \citet[e.g.,][]{SM02},
so we here only snow the results.  
The adsorption rate coefficient for electrons with the velocity distribution function 
given by Equation~\eqref{eq:fe} can be written as \citep{S41,SM02}
\beq
K_{de} = \pi a^2 \brac{v_e}\exp\left(-\frac{e^2|Z|}{a \kB T_e}\right),
\label{eq:Kde}
\eeq
where $\brac{v_e}$ is the mean electron speed given by Equation~\eqref{eq:ve}.
Equation~\eqref{eq:Kde} assumes that the grains in the plasmas on average charge negatively,
i.e., $Z < 0$, which is true as long as photoelectric and secondary electron emission 
from the grains are negligible \citep[e.g.,][]{SM02}.
Under this approximation, the Coulomb reduction factor 
$C_e$ has a simple expression,
\beq
C_e \equiv \frac{K_{de}}{\pi a^2 \brac{v_e}} =  \exp\left(-\frac{e^2|Z|}{a \kB T_e}\right).
\label{eq:Ce}
\eeq
Note that $C_e < 0$.

The ion adsorption rate coefficient under the Maxwellian approximation 
can be written as \citep[][Equation~(46) of \citetalias{OI15}]{SM02} 
\beqn
K_{di} &=& \pi a^2 \Biggl[ \sqrt{\dfrac{2 \kB T_i}{\pi m_i}}
\exp\left(-\frac{m_i\brac{{\bm v}_{i}}^2}{2 \kB T_i} \right)  
\nonumber \\
&&+ |\brac{{\bm v}_{i}}| 
\left( 1 + \frac{\kB T_i + 2e^2|Z|/a }{m_i \brac{{\bm v}_{i}}^2} \right) {\rm erf}
\left( \sqrt{\frac{m_i}{2\kB T_i}}|\brac{{\bm v}_{i}}| \right) \Biggr],
\label{eq:Kdi}
\eeqn
where  ${\rm erfc}(x)$ is the complementary error function.
The Coulomb attraction factor $C_i$ can formally be defined as 
\beq
C_i \equiv \frac{K_{di}}{\pi a^2 \brac{v_i}},
\label{eq:Ci}
\eeq
with the mean ion speed $\brac{v_i}$ given by
\beqn
\brac{v_i} &=& \sqrt{\dfrac{2 \kB T_i}{\pi m_i}}
\exp\left(-\frac{m_i\brac{{\bm v}_{i}}^2}{2 \kB T_i} \right)  
\nonumber \\
&&+ |\brac{{\bm v}_{i}}| 
\left( 1 + \frac{\kB T_i  }{m_i \brac{{\bm v}_{i}}^2} \right) {\rm erf}
\left( \sqrt{\frac{m_i}{2\kB T_i}}|\brac{{\bm v}_{i}}| \right),
\label{eq:vi}
\eeqn
which follows from the fact that $K_{di}$ must have the form $K_{di} = \pi a^2 \brac{v_i}$ in the case of $Z=0$. 
For $E' \gg E_{{\rm crit},i}$, $\brac{v_i}$ and $C_i$ have simple forms \citep[][Equation~(47) of \citetalias{OI15}]{SM02} 
\beq
\brac{v_i} \approx |\brac{{\bm v}_{i}}|, \qquad C_i \approx 1 + \frac{2e^2|Z|}{a m_i \brac{{\bm v}_{i}}^2}.
\eeq

The gas-phase recombination rate coefficient is taken as 
\beq
K_{\rm rec} = 2.4\times 10^{-7}\pfrac{T_e}{300~{\rm K}}^{-0.69}~{\rm cm^3~s^{-1}},
\label{eq:Krec}
\eeq
which assumes that the dominant positive ion is HCO$^+$ \citep{GBJD88}.

\bibliography{myrefs_190429}

\end{document}